\documentstyle[12pt]{article}

\begin{document}
\setcounter{page}{1}

\pagestyle{plain}
\vspace{1cm}
\begin{center}
\Large{\bf Generalized Uncertainty Principle, Modified Dispersion
Relations and the Early Universe Thermodynamics }\\
\small
\vspace{1cm} {\bf Kourosh Nozari}\quad and \quad {\bf Behnaz Fazlpour}\\
\vspace{0.5cm} {\it Department of Physics,
Faculty of Basic Sciences,\\
University of Mazandaran,\\
P. O. Box 47416-1467,
Babolsar, IRAN\\
e-mail: knozari@umz.ac.ir}

\end{center}
\vspace{1.5cm}

\begin{abstract}
In this paper, we study the effects of Generalized Uncertainty
Principle(GUP) and Modified Dispersion Relations(MDRs) on the
thermodynamics of ultra-relativistic particles in early universe. We
show that limitations imposed by GUP and particle horizon on the
measurement processes, lead to certain modifications of early
universe thermodynamics.
\end{abstract}
\section{Introduction}
Generalized Uncertainty Principle is a common feature of all
promising candidates of quantum gravity. String theory, loop quantum
gravity and noncommutative geometry(with deeper insight to the
nature of spacetime at Planck scale), all indicate the modification
of standard Heisenberg principle [1-10]. Recently it has been
indicated that within quantum gravity scenarios, a modification of
dispersion relation(relation between energy and momentum of a given
particle) is unavoidable[11-13]. There are some conceptual relations
between GUP and MDRs. These possible relations have been studied
recently[14,15]. These quantum gravity effects, in spite of being
small, are important since they can modify experimental results.
There are several efforts to provide experimental evidence of these
small effects. For example, Amelino-Camelia {\it et al}, by
investigation of potential sensitivity of Gamma-Ray Burster
observations to wave dispersion in vacuo, have outlined aspects of
an observational programme that could address possible detection of
these quantum gravity effects[16]. Amelino-Camelia and Piran have
argued that Planck-scale deformation of Lorentz symmetry can be a
solution to the Ultra High Energy Cosmic Rays(UHECR) with energies
above the GZK threshold and the TeV-$\gamma$ paradoxes[17]. Gambini
and Pullin have studied light propagation in the picture of
semi-classical spacetime that emerges in canonical quantum gravity
in the loop representation[18]. They have argued that in such a
picture, where space-time exhibits a polymer-like structure at
microscales, it is natural to expect departures from the perfect
non-dispersiveness of ordinary vacuum. They have evaluated these
departures by computing the modifications to Maxwell's equations due
to quantum gravity, and showing that under certain circumstances,
non-vanishing corrections appear that depend on the helicity of
propagating waves. These effects could lead to observable
cosmological predictions of the discrete nature of quantum
spacetime. Then, they have addressed to observations of
non-dispersiveness in the spectra of gamma-ray bursts at various
energies to constrain the type of semi-classical state that
describes the universe. Jacobson {\it et al} have shown that
threshold effects and Planck scale Lorentz violation are combined
constraints from high energy astrophysics[19]. These literatures
provide possible experimental schemes for detection of small quantum
gravity effects. However, there are two extreme domains: black hole
structure and early stages of the universe evolution where these
quantum gravity effects are dominant. Corrections to black hole
thermodynamics due to quantum gravitational effects of minimal
length and GUP have been studied extensively(see [20] and references
therein). On the other hand, part of the thermodynamical
implications of GUP and MDR have been studied by Amelino-Camelia
{\it et al}[21] and Nozari {\it et al}[22]. Thermodynamics of early
universe within standard Heisenberg principle has been studied by
Rahvar {\it et al}[23]. Since quantum gravitational effects are very
important in early stages of the universe evolution, it is natural
to investigate early universe thermodynamics within GUP and MDRs
frameworks. Here we are going to formulate thermodynamics of
ultra-relativistic particles in early universe within GUP and MDRs
frameworks. In the first step, using GUP as our primary input, we
calculate thermodynamical properties of ultra-relativistic particles
in early universe. In formulation of the early universe
thermodynamics within GUP framework, due to limitations imposed on
the measurement processes, two main points should be considered:
first due to casual structure of spacetime, maximum distance for
causal relation is particle horizon radius and secondly, there is a
minimum momentum imposed by GUP which restricts the minimum value of
energy. In the next step, for a general gaseous system composed of
ultra-relativistic particles, we find density of states using MDRs
with Bose-Einstein or Fermi-Dirac statistics and then thermodynamics
of the system will be followed. In each step we discuss ordinary
limits of our equations and we compare consequences of two
approaches.

\section{Preliminaries}
Emergence of the generalized uncertainty principle can be motivated
and finds support in the direct analysis of any quantum gravity
scenario. This means that GUP itself is a model independent concept.
Generally, GUP can be written as[24]
\begin{equation}
\delta x\delta p\geq\frac{\hbar}{2}\Big(1+\kappa(\delta
x)^2+\eta(\delta p)^2+\gamma\Big),
\end{equation}
where $\kappa$, $\eta$ and $\gamma$ are positive and independent of
$\delta x$  and $\delta p$ (but may in general depend on the
expectation values of $x$ and $p$). This GUP leads to a nonzero
minimal uncertainty in both position and momentum for positive
$\kappa$ and $\eta$[24]. If we set $\kappa=0$ we find
\begin{equation}
\delta x\delta p\geq\frac{\hbar}{2}\Big(1+\eta(\delta
p)^2+\gamma\Big).
\end{equation}
Since we are going to deal with absolutely smallest uncertainties,
we set $\gamma=0$ from now on. So we find
\begin{equation}
\delta x\delta p\geq\frac{\hbar}{2}(1+\eta(\delta p)^2).
\end{equation}
This relation leads to a nonzero minimal observable length of the
order of Planck length, $(\delta x)_{min}=\hbar\sqrt{\eta}$. Any
position measurement in quantum gravity has at least $(\delta
x)_{min}$ as its lower limit of position uncertainty. This relation
has an immediate consequence for the rest of statistical mechanics:
it modifies the {\it fundamental volume $\omega_{0}$} of accessible
phase space for representative points. In ordinary statistical
mechanics, it is impossible to define the position of a
representative point in the phase space of the given system more
accurately than the situation which is given by $(\delta q\, \delta
p)_{min}\geq \hbar$. In another words, around any point $(q,p)$ of
the (two dimensional) phase space, there exists an area of the order
$\hbar$ which the position of the representative point cannot be
pin-pointed. In ordinary statistical mechanics we have the following
definition of fundamental volume
\begin{equation}
\omega_0=(\delta q\,\delta p)^{3N}.
\end{equation}
Since in quantum gravity era $\delta p \sim p$, we can interpret
equation (3) as a generalization of $\hbar$,
$$ \hbar_{eff}=\hbar\Big(1+\eta p^2).$$ Therefore, we find
the following generalization of the fundamental volume
\begin{equation}
{(\omega_0)}_{eff}=[\hbar(1+\eta p^2)]^{3N}\equiv
(\hbar_{eff})^{3N}.
\end{equation}
Since the total number of microstates is given by
$\Omega=\frac{\omega}{{(\omega_0)}_{eff}}$ (here $\omega$ is the
volume of the accessible phase space), we see that GUP leads to a
reduction of accessible microstates and therefore a reduction of
entropy. In other words, when we approaches Planck scale regime with
high energy and momentum particles, the volume of the fundamental
cell increases in such away that eventually the number of
microstates tends to unity and therefore entropy vanishes. This is a
novel prediction of quantum gravity. Recently we have calculated
microcanonical entropy of an ideal gaseous system and we have
observed an unusual thermodynamics of systems in very
short distances or equivalently very high energy regime[22].\\
Another consequence of GUP in the form of relation (3), has been
formulated by Kempf {\it et al}[24]. They have shown that within the
momentum representation, the generalization of the scalar products
reads
\begin{equation}
\langle\psi|\phi\rangle=\int\frac{dp}{1+\eta p^2}\psi^{*}(p)\phi(p),
\end{equation}
where $\phi$ and $\psi$ are momentum space state functions. For
ultra-relativistic particles with $E=pc$, we should consider the
following generalization
\begin{equation}
dE\longrightarrow\frac{dE}{1+\eta E^{2}},
\end{equation}
where we have set $c=1$.\\
On the other hand, if we set $\eta=0$ in (1), we find
\begin{equation}
\label{math:1.2} \delta x\delta
p\geq\frac{\hbar}{2}\Big(1+\kappa(\delta x)^2+\gamma\Big),
\end{equation}
where for positive $\kappa$ leads to nonzero minimal uncertainty in
momentum. This statement leads to a space-dependent generalization
of $\hbar$. This type of generalization has nothing to do with
dynamics and there is no explicit physical interpretation of it at
least up to now.\\
From another perspective, in scenarios which consider spacetime foam
intuition in the study of quantum gravity phenomena, emergence of
modified dispersion relations takes place naturally[25]. As a
consequence, wave dispersion in the spacetime foam might resemble
wave dispersion in other media. Since Planck length fundamentally
set the minimum allowed value for wavelengths, a modified dispersion
relation can also be favored. Recently it has been shown that a
modified energy-momentum dispersion relation can also be introduced
as an observer-independent law[26]. In this case, the Planckian
minimum-wavelength hypothesis can be introduced as a physical law
valid in every frame. Therefore, the analysis of some
quantum-gravity scenarios has shown some explicit mechanisms for the
emergence of modified dispersion relations. For example, in the
framework of noncommutative geometry and loop quantum gravity
approaches this modified dispersion relations have been
motivated(see for example[21] and references therein). In most cases
one is led to consider a dispersion relation of the type(note that
from now on we set $c=\hbar=k_{B}=1$)
\begin{equation}
\label{math:1.5}
 (\vec{p})^2 = f(E,m;l_p) \simeq E^2 - \mu^2+ \alpha_1 l_p E^3
+ \alpha_2 l_p^2 E^4+ O\left(l_p^3 E^5 \right)
\end{equation}
where $f$ is the function that gives the exact dispersion relation,
and on the right-hand side we have assumed the applicability of a
Taylor-series expansion for $E \ll 1/l_p$. The coefficients
$\alpha_i$ can take different values in different quantum-gravity
proposals. Note that $m$ is the rest energy of the particle and the
mass parameter $\mu$ on the right-hand side is directly related to
the rest energy, but $\mu \neq m$ if the $\alpha_i$ do not all
vanish. Since we are working in Planck regime where the rest mass is
much smaller than the particle kinetic energy, there is no risk of
confusing between $m$ and $\mu$. While in the parametrization of (3)
we have included a possible correction term suppressed only by one
power of the Planck length, in GUP such a linear-in-$l_p$ term is
assumed not to be present. For the MDR a large number of alternative
formulations, including some with the linear-in-$l_p$ term, are
being considered, as they find support in different approaches to
the quantum-gravity problem, whereas all the discussions of a GUP
assume that the leading-order correction should be proportional to
the square of $l_p$ (as has been indicated by Amelino-Camelia {\it
et al}[21], linear-in-$l_p$ term in MDR has no support in string
theory analysis of black holes entropy-area relation and therefore
it seems that this term should not be present in MDR. Recently we
have shown that coefficients of all odd power of $E$ in MDR should
be zero[15]).\\
Within quantum field theory, the relation between particle
localization and its energy is given by $E\ge\frac{1}{\delta x}$,
where $\delta x$ is particle position uncertainty. It is obvious
that due to both GUP and  MDR this relation should be modified. In a
simple analysis based on the familiar derivation of the relation
$E\ge\frac{1}{\delta x}$[27], one can obtain the corresponding
generalized relation. Since we need this generalization in
forthcoming arguments, we give a brief outline of its derivation
here. We focus on the case of a particle of mass $M$ at rest, whose
position is being measured by a procedure involving a collision with
a photon of energy $E_\gamma$ and momentum $p_\gamma$. According to
Heisenberg's uncertainty principle, in order to measure the particle
position with precision $\delta x$, one should use a photon with
momentum uncertainty $\delta p_\gamma\ge\frac{1}{\delta x}$.
Following the standard argument[28], one takes this $\delta
p_\gamma\ge\frac{1}{\delta x}$ relation and converts it into the
relation $\delta E_\gamma\ge\frac{1}{\delta x}$ using the special
relativistic dispersion relation. Finally $\delta
E_\gamma\ge\frac{1}{\delta x}$ is converted into the relation
$M\ge\frac{1}{\delta x}$ because the measurement procedure requires
$\delta E \le M$, in order to ensure that the relevant energy
uncertainties are not large enough to allow the production of
additional copies of the particle whose position is being measured.
If indeed our quantum-gravity scenario hosts a Planck-scale
modification of the dispersion relation of the form (9) then clearly
the relation between $\delta p_\gamma$ and $\delta E_\gamma$ should
be re-written as follows
\begin{equation}
\label{math:1.6} \delta p_\gamma \simeq  \Big[1+ \alpha_1 l_p
E+3\big(\frac{\alpha_2}{2}-\frac{\alpha_1^2}{8}\big) l_p^2
E^2\Big]\delta E_\gamma.
\end{equation}
This relation will modify density of states for statistical systems.
Note that one can use GUP to find such relation between $\delta
p_\gamma$ and $\delta E_\gamma$[15].

\section{GUP and Early Universe Thermodynamics} Now we are going to
calculate thermodynamical properties of ultra-relativistic particles
in early universe, using the generalized uncertainty principle. We
consider the following GUP as our primary input,
\begin{equation}
\label{math:1.1} \delta x \delta p \geq
\pi\Bigg(1+\xi^{2}\frac{(\delta x)^{2}}{l_{p}^{2}}\Bigg),
\end{equation}
where $\xi$ is a dimensionless constant. Consider the early stages
of the universe evolution. Analogue to a particle inside a box, in
the case of the early universe one can consider a causal box (i.e.
particle horizon) which any observer in the universe has to do
measurements within this scale[29]. In the language of wave
mechanics, if $\Psi$ denotes the wave function of a given particle,
the probability of finding this particle by an observer outside its
horizon is zero, i. e. $|\Psi(x>horizon)|^{2}=0$. From the theory of
relativity, measurement of a stick length can be done by sending
simultaneous signals to the observer from the two endpoints, where
for the scales larger than the causal size, those signals need more
than the age of the universe to be received. Looking back to the
history of the universe, the particle horizon after the Planck era
grows as $H^{-1}$, but inflates to a huge size by the beginning of
inflationary epoch. Here $H$ is Hubble parameter. In the
pre-inflationary epoch, the maximum uncertainty in the location of a
particle, $\delta x =H^{-1}$ results in an uncertainty in the
momentum of the particle which is given by
\begin{equation}
\label{math:1.1}\delta p \geq\frac{\pi\xi^{2}}{ l_{p}^{2}H}+\pi H .
\end{equation}
This leads to a minimum uncertainty in momentum as
\begin{equation}
\label{math:1.1}(\delta p)_{min} = \frac{\pi\xi^{2}}{
l_{p}^{2}H}+\pi H.
\end{equation}
Therefore, we can conclude that(assuming that $p\sim \delta p$)
\begin{equation}
\label{math:1.1}p_{min} = \frac{\pi\xi^{2}}{ l_{p}^{2}H}+\pi H,
\end{equation}
which leads to
\begin{equation}
\label{math:1.1}E_{min} = \sqrt{3}\Big(\frac{\pi\xi^{2}}{
l_{p}^{2}H}+\pi H\Big),
\end{equation}
for ultra-relativistic particles in three space dimensions. Now,
suppose that
\begin{equation}
\label{math:1.1}E_{n}=n\vartheta,
\end{equation}
where $\vartheta$ is given by
\begin{equation}
\label{math:1.1}\vartheta=\frac{\pi\xi^{2} }{ l_{p}^{2}H}+\pi H.
\end{equation}
To obtain complete thermodynamics of the system, we calculate
partition function of the system and then we use standard
thermodynamical relations. In classical statistical mechanics,
partition function for a system composed of ultra-relativistic
noninteracting monatomic particles (Fermions or Bosons) is given by
\begin{equation}
\ln Z =\pm g\int_{0}^{\infty}\frac{4\pi n^{2}}{8}\ln(1\pm e^{-\beta
E_{n}})dn.
\end{equation}
In our case, due to limitation imposed by GUP and particle horizon,
we should consider the following generalization
\begin{equation}
\ln Z =\pm \frac{g\pi}{2\vartheta^3}\int_{E_{min}}^{\infty}\frac{
E^{2}}{1+ \eta E^{2}}\ln(1\pm e^{-\beta E})dE,
\end{equation}
where we have used relations (7), (15) and (16) respectively. By
definition, the entropy of the system is given by
\begin{equation}
S=-\frac{1}{V}\frac{\partial F}{\partial T},
\end{equation}
where $F$ is the free energy of the system defined as
\begin{equation}
F=-\frac{1}{\beta}\ln Z.
\end{equation}
So the entropy of the system can be written as
\begin{equation}
S=\frac{1}{V}\Bigg[\pm
\frac{g\pi}{2\vartheta^3}\int_{E_{min}}^{\infty}\frac{ E^{2}}{1+
\eta E^{2}}\ln(1\pm e^{-\beta
E})dE+\frac{g\pi\beta}{2\vartheta^3}\int_{E_{min}}^{\infty}\frac{
E^{3}}{1+ \eta E^{2}}\frac{dE}{e^{\beta E}\pm1}\Bigg].
\end{equation}
For ultra-relativistic fermions this relation leads to the following
expression
$$S_{f}=\frac{g}{2\pi^{2}(1+D)}\Bigg[\frac{7}{90}\frac{\pi^4}{\beta^3}-
 \frac{31}{210}\frac{\eta\pi^6}{\beta^5}+\frac{1016}{1680}\frac{\eta^{2}\pi^{8}}{\beta^{7}}
 -E_{min}^3\Bigg(\frac{1}{3}-\frac{\eta E_{min}^{2}}{5}+\frac{\eta^{2} E_{min}^{4}}
 {7}\Bigg)\ln(1+e^{-\beta E_{min}})$$
\begin{equation}
 -\frac{4}{3}\beta I_{3}+\frac{6}{5}\eta\beta I_{5} -\frac{8}{7}\eta^{2}\beta I_{7}+... \Bigg],
\end{equation}
where for simplicity we have defined
$$I_{j}=\int_{0}^{E_{min}}
 \frac {E^j dE}{e^{\beta E}+1}.$$
While for bosons we find
$$S_{b}=\frac{g}{2\pi^{2}(1+D)}\Bigg[\frac{4}{45}\frac{\pi^4}{\beta^3}-
 \frac{48}{315}\frac{\eta\pi^6}{\beta^5}+\frac{64}{105}\frac{\eta^{2}\pi^{8}}{\beta^{7}}
 +E_{min}^3\Bigg(\frac{1}{3}-\frac{\eta E_{min}^{2}}{5}+\frac{\eta^{2} E_{min}^{4}}
 {7}\Bigg)\ln(1-e^{-\beta E_{min}})$$
\begin{equation}
-\frac{4}{3}\beta J_{3}+\frac{6}{5}\eta \beta J_{5}-
\frac{8}{7}\eta^{2}\beta J_{7}+...  \Bigg],
\end{equation}
where $$J_{j}=\int_{0}^{E_{min}}
 \frac {E^j dE}{e^{\beta E}-1}.$$
In these equations $D$ is defined as
$$D=(\frac{A^{3}}{B^{3}}+3\frac{A^{2}}{B^{2}}+3\frac{A}{B})\quad \quad and\quad
\vartheta=A+B$$ with $A=\frac{\pi\xi^{2}}{ l_{p}^{2}H}$ and $B=\pi
H$. Note that both of the equations (23) and (24) are well behavior
in high and low temperature limits. In the standard situation, we
have $\xi=0$, $\eta=0$ and $E_{min}=0$. So we find the well-known
and standard results for entropy of the corresponding
ultra-relativistic fermionic or bosonic systems. From (22) we find
the following expression for standard entropy
\begin{equation}
S=\frac{4}{3}\frac{\beta g}{2\pi^2}\int_{0}^{\infty}\frac{E^3
dE}{e^{\beta E}\pm 1},
\end{equation}
which leads to
\begin{equation}
S_{f}=\frac{ g}{2\pi^2}\frac{7}{90}\frac{\pi^4}{\beta^3},
\end{equation}
and
\begin{equation}
S_{b}=\frac{ g}{2\pi^2}\frac{4}{45}\frac{\pi^4}{\beta^3},
\end{equation}
for fermions and bosons respectively.\\
Now the pressure of the ultra-relativistic gas is given by
$P=\frac{1}{\beta V}\ln Z$. For fermions and bosons we find
respectively
$$P_{f}=\frac{g}{2\pi^{2}(1+D)\beta}\Bigg[\frac{7}{360}
\frac{\pi^4}{\beta^3}-\frac{31}{1260}\frac{\eta \pi^6}{\beta^5}
+\frac{127}{1680}\frac{\eta^{2}
\pi^8}{\beta^7}-E_{min}^{3}\Bigg(\frac{1}{3} -\eta
\frac{E_{min}^2}{5}+\eta^{2}\frac{E_{min}^{4}}{7}
\Bigg)\ln(1+e^{-\beta E_{min}})$$
\begin{equation}
-\frac{\beta}{3}I_{3}+\frac{\beta}{5}\eta
I_{5}-\frac{\beta}{7}\eta^{2}I_{7}+... \Bigg],
\end{equation}
and
$$P_{b}=\frac{g}{2\pi^{2}(1+D)\beta}\Bigg[\frac{1}{45}
\frac{\pi^4}{\beta^3}-\frac{8}{315}\frac{\eta \pi^6}{\beta^5}
+\frac{8}{105}\frac{\eta^{2}
\pi^8}{\beta^7}+E_{min}^{3}\Bigg(\frac{1}{3} -\eta
\frac{E_{min}^2}{5}+\eta^{2}\frac{E_{min}^{4}}{7}
\Bigg)\ln(1-e^{-\beta E_{min}})$$
\begin{equation}
-\frac{\beta}{3} J_{3}+\frac{\beta}{5}\eta
J_{5}-\frac{\beta}{7}\eta^{2} J_{7}+... \Bigg].
\end{equation}
In the standard situation, we find the following well-known result
\begin{equation}
P=\frac{ g}{6\pi^2}\int_{0}^{\infty}\frac{E^3 dE}{e^{\beta E}\pm 1},
\end{equation}
which leads to
\begin{equation}
P_{f}=\frac{ g}{2\pi^2}\frac{7}{360}\frac{\pi^4}{\beta^4},
\end{equation}
and
\begin{equation}
P_{b}=\frac{ g}{2\pi^2}\frac{1}{45}\frac{\pi^4}{\beta^4},
\end{equation}
for fermions and bosons respectively. \\
The specific heat of the system which is defined as
\begin{equation}
C_{V}=T \Bigg(\frac{\partial S}{\partial T}\Bigg)_{V},
\end{equation}
can be written in the following closed form
\begin{equation}
C_{V}=\frac{g\beta^2}{2\pi^{2}(1+D)}\int_{E_{min}}^{\infty}\frac{E^4}{1+\eta
E^2} \frac{dE}{e^{-\beta E}\big(e^{\beta E}\pm1)^{2}}.
\end{equation}
One can obtain explicit form of $C_{V}$ for fermions and bosons
using relation (33), (23) and (24). A simple calculation gives
$$C_{Vf}=\frac{g}{2\pi^{2}(1+D)}\Bigg[\frac{21}{90}
\frac{\pi^4}{\beta^3}-\frac{155}{210}\frac{\eta\pi^6}{\beta^5}+
\frac{7112}{1680} \frac{\eta^{2}\pi^{8}}{\beta^{7}}
 -\frac{\beta E_{min}^4}{e^{\beta E_{min}}+1}\Bigg(\frac{1}{3}-
 \frac{\eta E_{min}^{2}}{5}+\frac{\eta^{2} E_{min}^{4}}
 {7}\Bigg)$$
\begin{equation}
 +\frac{4}{3}\beta I_{3}-\frac{6}{5}\eta\beta I_{5} +\frac{8}{7}\eta^{2}\beta I_{7}
 -\frac{4}{3}\frac{dI_3}{dT}+\frac{6}{5}\eta\frac{dI_5}{dT}-\frac{8}{7}\eta^2
 \frac{dI_7}{dT}+... \Bigg],
\end{equation}
and
$$C_{Vb}=\frac{g}{2\pi^{2}(1+D)}\Bigg[\frac{12}{45}\frac{\pi^4}{\beta^3}-
\frac{240}{315}\frac{\eta\pi^6}{\beta^5}+\frac{448}{105}\frac{\eta^{2}\pi^{8}}{\beta^{7}}
 +\frac{\beta E_{min}^4}{e^{\beta E_{min}}-1}\Bigg(\frac{1}{3}-\frac{\eta E_{min}^{2}}{5}+\frac{\eta^{2} E_{min}^{4}}
 {7}\Bigg)$$
\begin{equation}
+\frac{4}{3}\beta J_{3}-\frac{6}{5}\eta \beta J_{5}+
\frac{8}{7}\eta^{2}\beta
J_{7}-\frac{4}{3}\frac{dJ_3}{dT}+\frac{6}{5}\eta\frac{dJ_5}{dT}-\frac{8}{7}
\eta^2\frac{dJ_7}{dT}+... \Bigg],
\end{equation}
for specific heat of fermions and bosons respectively. In the
standard case we find
\begin{equation}
C_{Vf}=3\times\frac{g}{2\pi^{2}}\frac{7}{90}\frac{\pi^4}{\beta^3}=3S_{f},
\end{equation}
and
\begin{equation}
C_{Vb}=3\times\frac{g}{2\pi^{2}}\frac{4}{45}\frac{\pi^4}{\beta^3}=3S_{b}.
\end{equation}
Figure 1 shows the values of entropy in different situations. In
standard thermodynamics of ultra-relativistic fermionic or bosonic
gas, the entropy of the system tends to zero in $T_{0}=0$. This
situation is shown in Figure 1, (a) and (b). Within GUP framework,
entropy tends to zero in a nonzero temperature, that is, for
$T>T_{0}$. This is a result of quantum fluctuation of spacetime
itself. Figure 2 shows the corresponding behavior of pressure as a
function of temperature. Note that these figures are plotted in
arbitrary units and they show only general behaviors of the
functions. Figure 3 shows the behavior of specific heat of the
system in various conditions. In GUP framework, the general behavior
of $C_V$ has considerable departure from its standard counterpart in
high temperature regime.

\section{MDR and Early Universe Thermodynamics}
Now we are going to formulate early universe thermodynamics within
MDR framework. We consider a gaseous system composed of
ultra-relativistic monatomic, non-interacting particles. First we
derive the density of states. Consider a cubical box with edges of
length $L$ (and volume $V=L^3$) consisting black body
radiation(photons). The wavelengths of the photons are subject to
the boundary condition $\frac{1}{\lambda}=\frac{n}{2L}$, where $n$
is a positive integer. This condition implies, assuming that the de
Broglie relation is left unchanged, that the photons have
(space-)momenta that take values $p=\frac{n}{2L}$. Thus momentum
space is divided into cells of volume
$V_p=\left(\frac{1}{2L}\right)^3=\frac{1}{8V}$. From this point, it
follows that the number of modes with momentum in the interval
$[p,p+dp]$ is given by
\begin{equation}
g(p) dp =8\pi V p^2 dp
\end{equation}
Assuming a MDR of the type parameterized in (9) one then finds that
($m=0$ for photons)
\begin{equation}
p\simeq E \Big[1+\frac{\alpha_1}{2} l_p
E+\big(\frac{\alpha_2}{2}-\frac{\alpha_1^2}{8}\big) l_p^2 E^2\Big]
\end{equation}
and
\begin{equation}
dp \simeq  \Big[1+ \alpha_1 l_p
E+\big(\frac{3}{2}\alpha_2-\frac{3}{8}\alpha_1^2\big) l_p^2 E^2\Big]
dE
\end{equation}
Using this relation in (39), one obtains
\begin{equation}
g(E) dE = 8\pi V \left[1+2 \alpha_1 l_p E +
5\Big(\frac{1}{2}\alpha_{2}+\frac{1}{8}\alpha_{1}^{2}\Big)l_{p}^{2}
E^2\right] E^2 dE.
\end{equation}
This is density of states which we use in our calculations. Note
that we have not set $\alpha_{1} =0$ to ensure generality of our
discussions, but we will discuss corresponding situation at the end
of our calculations.\\
To obtain thermodynamics of the system under consideration, we start
with the partition function of fermions and bosons,
\begin{equation}
\label{math:1.1} \ln Z =\pm\int_{E_{min}}^{\infty}g(E)\ln(1\pm
e^{-\beta E})dE,
\end{equation}
where $+$ and $-$ stand for fermions and bosons respectively and
$\beta=1/T$ since $k_{B}=1$. Using equation(42) in the following
form
\begin{equation}
\label{math:1.1} g(E)dE=8 \pi V(1+a E +b E^{2})E^{2}dE,
\end{equation}
where for simplicity we have defined\, $ a=2\alpha_{1}l_{p}$\,\, and
$\,\,b=5\Big(\frac{1}{2}\alpha_{2}+\frac{1}{8}\alpha_{1}^{2}\Big)l_{p}^{2}$,\,\,
one can compute the integral of equation(43) to find the following
expression for entropy of fermions and bosons
\begin{equation}
 S =\pm\frac{1}{V}\int_{E_{min}}^{\infty}g(E)\ln(1\pm e^{-\beta E})dE+\frac{
 \beta}{V}
 \int_{E_{min}}^{\infty}\frac{g(E)E dE}{e^{\beta E}\pm1}.
\end{equation}
This relation can be written as follows
$$S=\pm\frac{1}{V}\int_{0}^{\infty}g(E)\ln(1\pm e^{-\beta
E})dE+\frac{ \beta}{V}
 \int_{0}^{\infty}\frac{g(E)E dE}{e^{\beta E}\pm1}$$
\begin{equation}
 \mp\frac{1}
 {V}\int_{0}^{E_{min}}g(E)\ln(1\pm e^{-\beta E})dE-\frac{
 \beta}{V}\int_{0}^{E_{min}}\frac{g(E)E dE}{e^{\beta E}\pm1}.
\end{equation}
By calculating this integral, we find for fermions and bosons
respectively
$$S_{f}=8\pi\Bigg[\frac{7}{90}\frac{\pi^4}{\beta^3}+\frac{225}
 {8}\frac{a\zeta(5)}{\beta^4}+\frac{31}{210}\frac{b\pi^6}{\beta^5}
 -E_{min}^3\Bigg(\frac{1}{3}+\frac{aE_{min}}{4}+\frac{bE_{min}^{2}}
 {5}\Bigg)\ln(1+e^{-\beta E_{min}})$$
\begin{equation}
 -\frac{4}{3}\beta I_{3}- \frac{5}{4}\beta a I_{4}-\frac{6}{5}\beta b I_{5} +...  \Bigg]
\end{equation}
and
$$S_{b}=8\pi\Bigg[\frac{4}{45}\frac{\pi^4}{\beta^3}+\frac{30 a\zeta(5)}{\beta^4}
 +\frac{48}{315}\frac{b\pi^6}{\beta^5}
 +E_{min}^3\Bigg(\frac{1}{3}+\frac{aE_{min}}{4}+\frac{bE_{min}^{2}}
 {5}\Bigg)\ln(1-e^{-\beta E_{min}})$$
\begin{equation}
-\frac{4}{3}\beta J_{3}- \frac{5}{4}\beta a J_{4}- \frac{6}{5}\beta
b J_{5}+... \Bigg].
\end{equation}
One may ask about the relation between these two results and
corresponding results of GUP, that is, relations (23) and (24).
Although these results seem to be different in their $\beta$
dependence, but note that if we set $\alpha_{1}=0$(which is
reasonable regarding the argument presented in page 5), we find
$a=0$ and then $\beta$ dependence of our findings will coincide with
each other. The only difference which remains is the differences
between numerical factors. This argument shows that essentially the
results of GUP and MDRs for thermodynamics of the early universe do
not differ with each other in their temperature dependence and
overall behaviors.\\
In the standard situation, we have $a=b=0$ and $E_{min}=0$, so we
find
\begin{equation}
S=\frac{4}{3}(8 \pi \beta)\int_{0}^{\infty}\frac{E^3 dE}{e^{\beta
E}\pm 1}.
\end{equation}
For entropy of fermions and bosons we find respectively
\begin{equation}
S_{f}=8 \pi \frac{7}{90}\frac{\pi^4}{\beta^3},
\end{equation}
and
\begin{equation}
S_{b}=8 \pi \frac{4}{45}\frac{\pi^4}{\beta^3}.
\end{equation}
In the presence of MDR, the pressure of corresponding systems are
$$P_{f}=\frac{8 \pi}{\beta}\Bigg[\frac{7}{360}
\frac{\pi^4}{\beta^3}+\frac{45}{8}\frac{a
\zeta(5)}{\beta^4}+\frac{31}{1260}\frac{b \pi^6}{\beta^5}
-E_{min}^{3}\Bigg(\frac{1}{3} +a \frac{E_{min}}{4}+b
\frac{E_{min}^{5}}{5} \Bigg)\ln(1+e^{-\beta E_{min}})$$
\begin{equation}
-\frac{\beta}{3}I_{3}-\frac{\beta}{4}a I_{4}-\frac{\beta}{5}b
I_{5}+... \Bigg],
\end{equation}
$$P_{b}=\frac{8 \pi}{\beta}\Bigg[\frac{1}{45}
\frac{\pi^4}{\beta^3}+6\frac{a\zeta(5)}{\beta^{4}}
+\frac{8}{315}\frac{b \pi^6}{\beta^5} +E_{min}^{3}\Bigg(\frac{1}{3}
+a \frac{E_{min}}{4}+b \frac{E_{min}^{5}}{5} \Bigg)ln(1-e^{-\beta
E_{min}})$$
\begin{equation}
-\frac{\beta}{3}J_{3}-\frac{\beta}{4}a J_{4}-\frac{\beta}{5}b
J_{5}+... \Bigg],
\end{equation}
for fermions and bosons respectively. In the standard situation we
find the following well-known relation
\begin{equation}
P=\frac{ 8 \pi}{3}\int_{0}^{\infty}\frac{E^3 dE}{e^{\beta E}\pm 1},
\end{equation}
which for fermions and bosons leads to
\begin{equation}
P_{f}=8 \pi \frac{7}{360}\frac{\pi^4}{\beta^4},
\end{equation}
and
\begin{equation}
P_{b}=8 \pi \frac{1}{45}\frac{\pi^4}{\beta^4}
\end{equation}
respectively.\\
The specific heat of the system can be written in the following
closed form
\begin{equation}
C_{V}=\frac{\beta^2}{V}\int_{E_{min}}^{\infty}\frac{g(E)E^2
dE}{e^{-\beta E}(e^{\beta E}\pm1)^{2}}.
\end{equation}
One can use relations (33), (47) and (48) to find the following
explicit results for fermions and bosons respectively
$$C_{Vf}=8\pi\Bigg[\frac{21}{90}\frac{\pi^4}{\beta^3}+\frac{900}
 {8}\frac{a\zeta(5)}{\beta^4}+\frac{155}{210}\frac{b\pi^6}{\beta^5}
 -\frac{\beta E_{min}^4}{e^{\beta E_{min}}+1}\Bigg(\frac{1}{3}+
 \frac{aE_{min}}{4}+\frac{bE_{min}^{2}}
 {5}\Bigg)$$
\begin{equation}
 +\frac{4}{3}\beta I_{3}+ \frac{5}{4}\beta a I_{4}+\frac{6}{5}\beta b I_{5}
 -\frac{4}{3}\frac{dI_3}{dT}-\frac{5}{4}a\frac{dI_4}{dT}-
 \frac{6}{5}b\frac{dI_5}{dT} +...  \Bigg],
\end{equation}
and
$$C_{Vb}=8\pi\Bigg[\frac{12}{45}\frac{\pi^4}{\beta^3}+\frac{120 a\zeta(5)}{\beta^4}
 +\frac{240}{315}\frac{b\pi^6}{\beta^5}
 +\frac{\beta E_{min}^4}{e^{\beta E_{min}}-1}\Bigg(\frac{1}{3}+\frac{aE_{min}}{4}+\frac{bE_{min}^{2}}
 {5}\Bigg)$$
\begin{equation}
 +\frac{4}{3}\beta J_{3}+ \frac{5}{4}\beta a J_{4}+\frac{6}{5}\beta b J_{5}
 -\frac{4}{3}\frac{dJ_3}{dT}-\frac{5}{4}a\frac{dJ_4}{dT}-\frac{6}{5}b
 \frac{dJ_5}{dT}+... \Bigg].
\end{equation}
In the standard case we find
\begin{equation}
C_{Vf}=3\times8\pi\frac{7}{90}\frac{\pi^4}{\beta^3}=3S_{f},
\end{equation}
and
\begin{equation}
C_{Vb}=3\times8\pi\frac{4}{45}\frac{\pi^4}{\beta^3}=3S_{b}
\end{equation}
respectively.\\
As has been indicated, there are severe constraints on the
functional form of MDR which these constraint are motivated when one
compares black hole entropy-area relation in different points of
view[15,21]. In this case we should set $\alpha_{1}=0$ which leads
to $a=0$. We find from (47) and (48) the following expressions for
entropy of fermions and bosons respectively
$$S_{f}=8\pi\Bigg[\frac{7}{90}\frac{\pi^4}{\beta^3}+\frac{31}{210}
\frac{b'\pi^6}{\beta^5}
 -E_{min}^3\Bigg(\frac{1}{3}++\frac{b'E_{min}^{2}}
 {5}\Bigg)\ln(1+e^{-\beta E_{min}})$$
\begin{equation}
 -\frac{4}{3}\beta I_{3}-\frac{6}{5}\beta b' I_{5} +...  \Bigg],
\end{equation}
and
$$S_{b}=8\pi\Bigg[\frac{4}{45}\frac{\pi^4}{\beta^3}+\frac{48}{315}
\frac{b'\pi^6}{\beta^5}
+E_{min}^3\Bigg(\frac{1}{3}+\frac{b'E_{min}^{2}}
{5}\Bigg)\ln(1-e^{-\beta E_{min}})$$
\begin{equation}
 -\frac{4}{3}\beta J_{3}-\frac{6}{5}\beta b' J_{5}+... \Bigg],
\end{equation}
where $b'=\frac{5}{2}\alpha_{2}l_{p}^2$. These statements for
partition function are more realistic since black hole
thermodynamics within MDRs when is compared with exact
solution of string theory, suggest the vanishing of $\alpha_{1}$.\\
It is important to note that the formalism presented in this section
is not restricted to early universe. Actually, it can be applied to
any statistical system composed of ultra-relativistic monatomic
noninteracting particles which has a minimum accessible energy .\\
The Possible relation between GUP and MDRs itself is under
investigation[14,15]. Generally these two features of quantum
gravity scenarios are not equivalent, but as Hossenfelder has shown,
they can be related to each other[14](see also [15]). As a result,
it is natural to expect that under special circumstances, our
results for early universe thermodynamics within GUP and MDRs should
transform to each other. This is a transformation between
coefficients of our equations and overall behaviors of
thermodynamical quantities, specially their temperature dependence
are similar.
\section{Summary}
GUP and MDRs have found strong supports from string theory,
noncommutative geometry and loop quantum gravity. There are many
implications, originated from GUP and MDRs, for the rest of the
physics. From a statistical mechanics point of view, GUP changes the
volume of the fundamental cell of the phase space in a momentum
dependent manner. On the other hand, MDR leads to a modification of
density of states. These quantum gravity features have novel
implications for statistical properties of thermodynamical systems.
Here we have studied thermodynamics of early universe within both
GUP and MDRs. We have considered early universe as a statistical
system composed of ultra-relativistic particles. Since both particle
horizon distance and GUP impose severe constraint on measurement
processes, the statistical mechanics of the system should be
modified to contain these constraint. Since GUP and MDRs are quantum
gravitation effects, the modified thermodynamics within GUP and MDRs
tends to standard thermodynamics in classical limits. There are
severe constraints on the functional form of MDRs from string theory
considerations. When we consider these constraints, the results of
MDRs and GUP for thermodynamics of early universe tends to each
other in their general temperature dependence and they differs only
in their numerical factors. This fact may be interpreted so that GUP
and MDRs essentially are not different concepts of quantum gravity
proposal. Although the exact relation between GUP and MDRs is not
known yet, our formalism of early universe shows the very close
relation between these two aspects of quantum gravity.\\
In standard statistical mechanics of bosonic  and fermionic gases,
the entropy of the system tends to zero in $T_{0}=0$. As our
equations and corresponding numerical result show, within GUP
framework entropy of the system tends to zero in a temperature
larger than zero( $T>T_{0}$). This is a consequence of the relation
(5). The volume of the fundamental cell of phase space increases due
to GUP. Note that MDRs give entropy-temperature relation which has
no difference with GUP result in its general behavior. Figure 2
shows the pressure of the system versus temperature. Pressure tends
to zero in a temperature larger than $T_{0}=0$. The same behavior is
repeated by specific heat of the system. So, our analysis shows an
unusual thermodynamics for statistical systems in quantum gravity
eras. This unusual behaviors have been seen in other context such as
black hole thermodynamics[30,31].

\begin{figure}[ht]
\begin{center}
\includegraphics{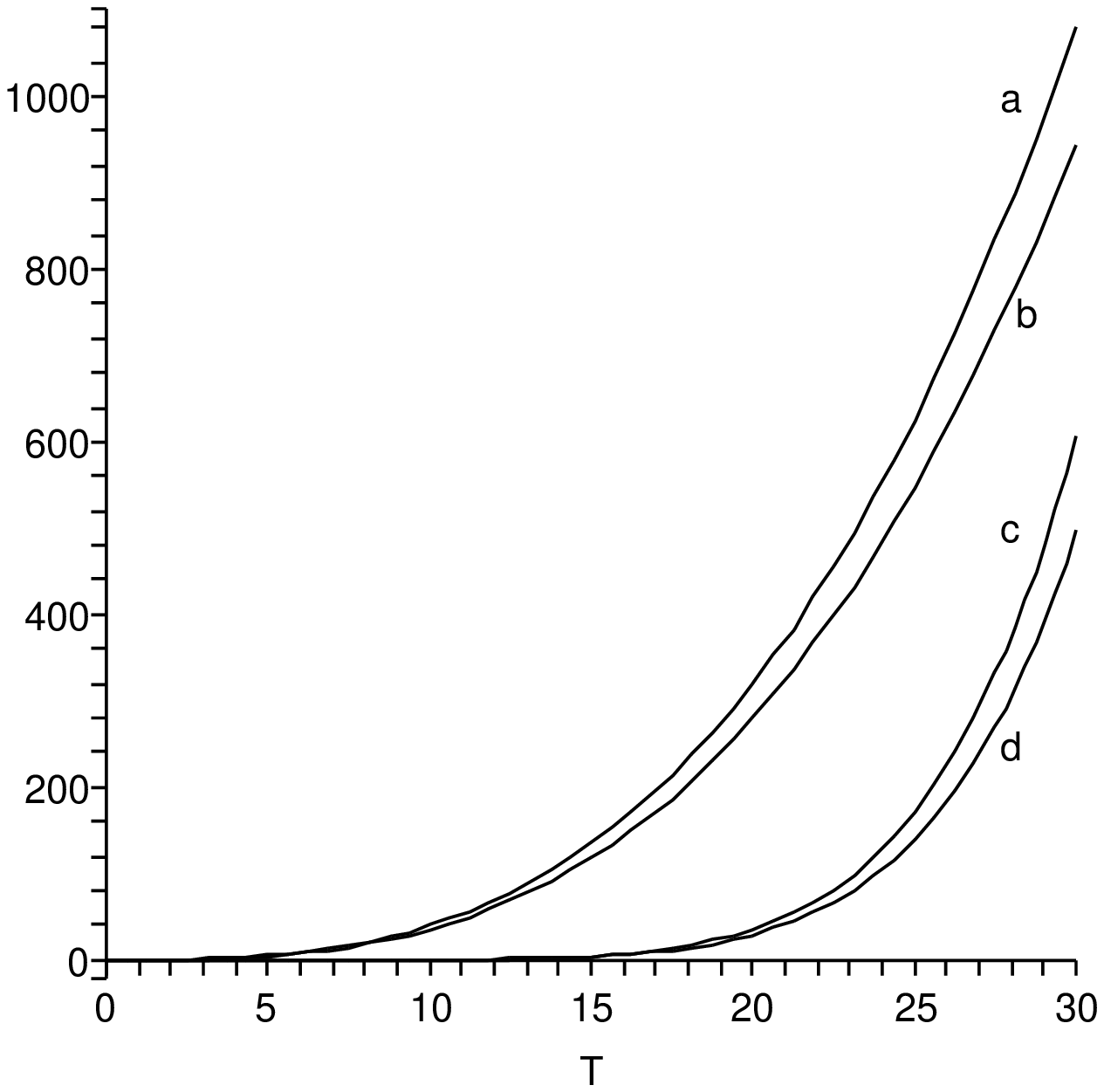}
\end{center}
\vspace{16 cm}
 \caption{\small {Entropy of ultra-relativistic monoatomic gaseous system for
 (a)standard bosonic gas (b) standard fermionic gas (c) bosonic
 gas within GUP and (d) fermionic gas within GUP.}}
 \label{fig:1}
 \end{figure}

 \begin{figure}[ht]
\begin{center}
\includegraphics{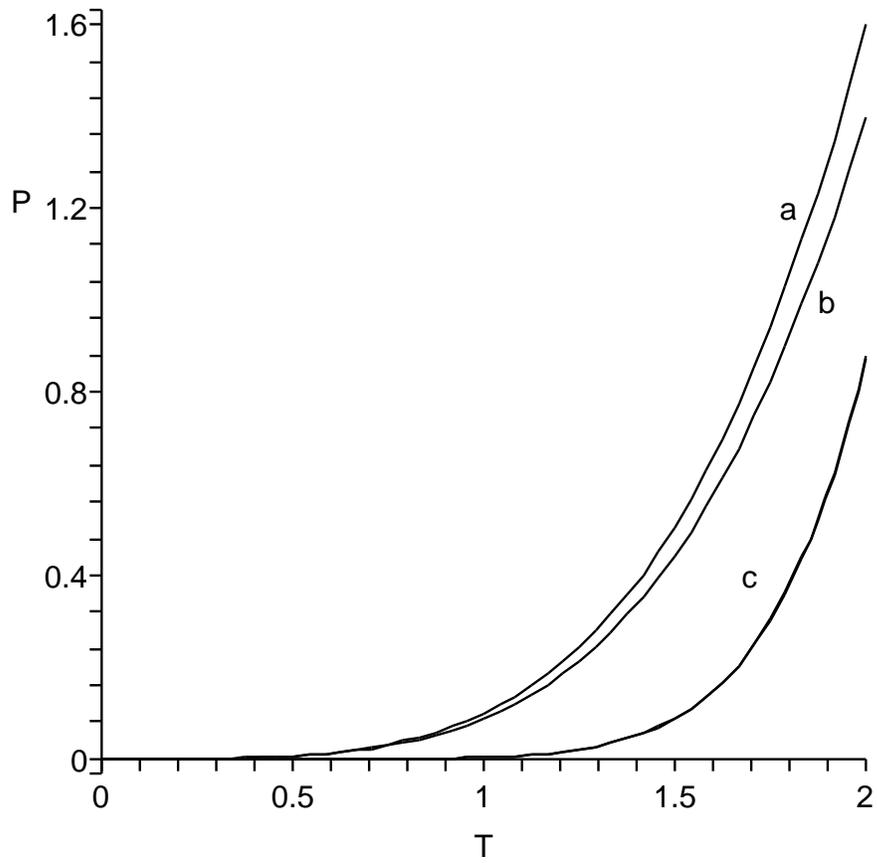}
\end{center}
\vspace{16 cm}
 \caption{\small {Pressure of ultra-relativistic monoatomic gaseous system for
 (a)standard bosonic gas (b) standard fermionic gas (c) bosonic
 and fermionic gas within GUP. The difference between bosonic and fermionic
 gasses in this case is not considerable.}}
 \label{fig:1}
 \end{figure}

 \begin{figure}[ht]
\begin{center}
\includegraphics{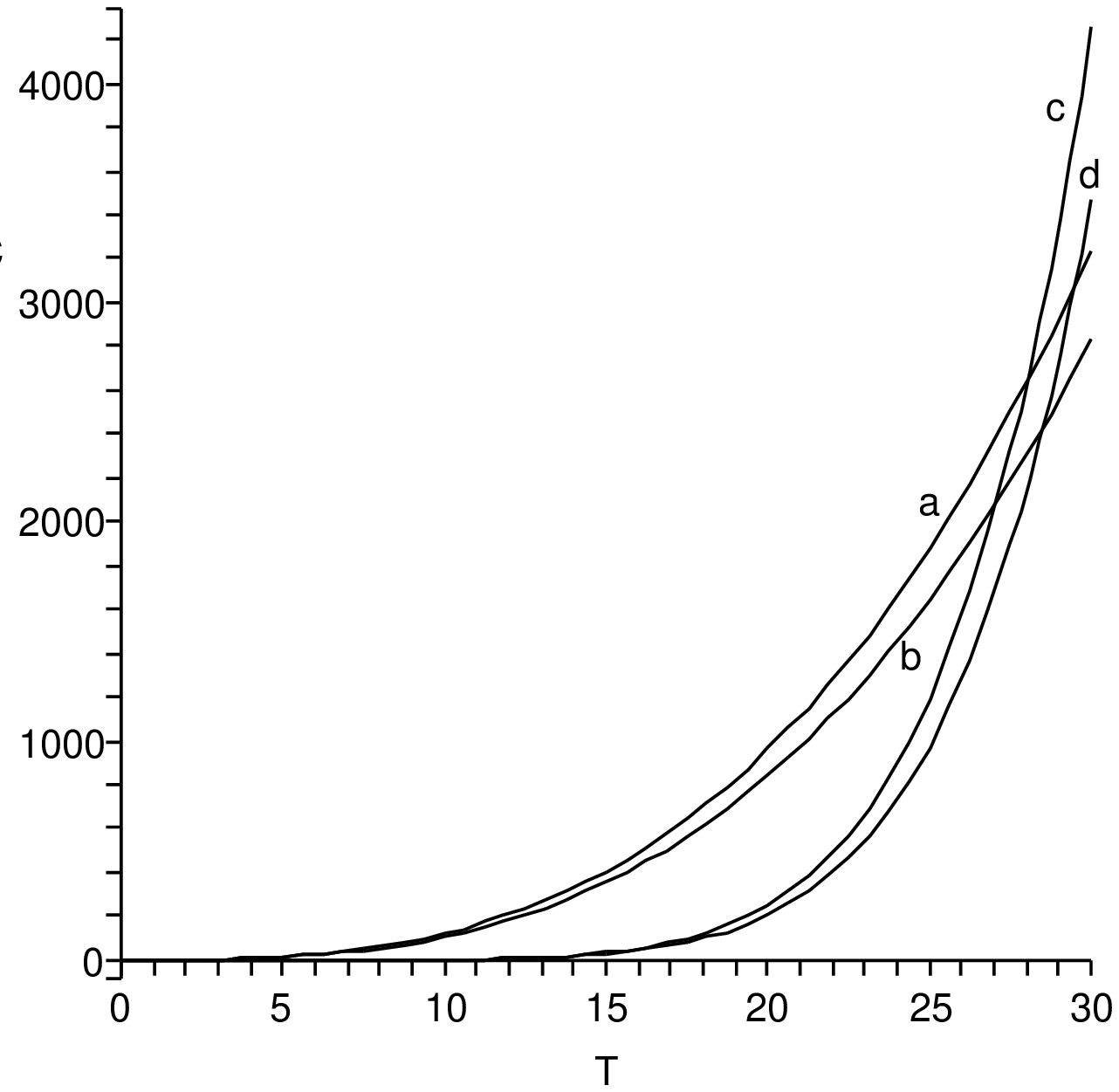}
\end{center}
\vspace{16 cm}
 \caption{\small {Heat Capacity of ultra-relativistic monoatomic gaseous system for
 (a)standard bosonic gas (b) standard fermionic  gas (c) bosonic gas
 within GUP  and (d) fermionic gas within GUP.}}
 \label{fig:1}
 \end{figure}

\end{document}